\documentclass[12pt]{article}
\usepackage[margin=1in]{geometry}
\usepackage[absolute,overlay]{textpos}  
\usepackage{authblk}
\usepackage{titlesec}
\usepackage{adjustbox}
\usepackage{graphicx}
\usepackage{url}
\usepackage{setspace}
\usepackage{hyperref}
\usepackage{parskip}
\usepackage{tcolorbox}
\usepackage{indentfirst}
\usepackage{soul}
\usepackage{tabularx}
\usepackage{hyperref}
\usepackage{multirow}
\usepackage{diagbox}
\usepackage{amsfonts}
\usepackage[numbers]{natbib}
\newcommand{\tickYes}{\checkmark}
\usepackage{pifont}
\newcommand{\tickNo}{\hspace{1pt}\ding{55}}
\usepackage{multicol}
\usepackage{vwcol}

\usepackage{subcaption}

\usepackage{lscape}
\definecolor{viol}{RGB}{161,36,137}
\usepackage{notoccite}
\usepackage{titlesec}
\usepackage{hyperref}
\usepackage{adjustbox}
\usepackage{float}
\usepackage{afterpage} 
\usepackage{placeins} 
\usepackage{tcolorbox} 
\usepackage{setspace} 
\usepackage[normalem]{ulem}

\usepackage{arydshln} 
\usepackage{array}
\usepackage{booktabs} 

\usepackage{tablefootnote}
\usepackage{setspace}

\usepackage{svg}
\svgsetup{inkscapelatex=false}

\newcounter{subsubsubsection}[subsubsection]
\renewcommand\thesubsubsubsection{\thesubsubsection.\arabic{subsubsubsection}}

\titleformat{\section}
  {\LARGE\bfseries} 
  {\thesection}{1em}{}

\titleformat{\subsection}
  {\large\bfseries} 
  {\thesubsection}{1em}{}

\titleformat{\subsubsection}
  {\normalsize\bfseries} 
  {\thesubsubsection}{1em}{}

\titleformat{\subsubsubsection}
  {\normalfont\normalsize\itshape}{\thesubsubsubsection}{1em}{}
\titlespacing{\subsubsubsection}
{0pt}{3.25ex plus 1ex minus .2ex}{1.5ex plus .2ex}

\titleformat{\section}{\normalfont\normalsize\bfseries}{\thesection}{1em}{}
\titleformat{\subsection}{\normalfont\small\bfseries}{\thesubsection}{1em}{}

\title{\textbf{\small RTGPU: Real-Time Computing with Graphics Processing Units}}

\author[1]{\small Atiyeh Gheibi-Fetrat}
\author[1]{\small Amirsaeed Ahmadi-Tonekaboni}
\author[1]{\small Farzam Koohi-Ronaghi}
\author[1]{\small Pariya Hajipour}
\author[1]{\small Sana Babayan-Vanestan}
\author[1]{\small Fatemeh Fotouhi}
\author[1]{\small Elahe Mortazavian-Farsani}
\author[1]{\small Pouria Khajehpour-Dezfouli}
\author[2]{\small Sepideh Safari}
\author[1]{\small Shaahin Hessabi}
\author[1,2]{\small Hamid Sarbazi-Azad}

\affil[1]{\small Sharif University of Technology, Tehran, Iran}
\affil[2]{\small Institute for Research in Fundamental Sciences (IPM), Tehran, Iran}

\date{}

\begin{document}
\maketitle

\begin{textblock*}{\textwidth}[0.5,1](105mm,275mm) 
\begin{tcolorbox}[colback=gray!10, colframe=black, boxrule=0.4pt, sharp corners, fontupper=\footnotesize, width=\textwidth]
\textbf{Note:} This document provides a concise summary of the book \textit{RTGPU}. Due to copyright 
restrictions, the full content is not reproduced here; readers are referred to the complete book for more 
comprehensive details. All listed authors contributed substantially to the development of the full publication. 
In compliance with publishing guidelines, undergraduate student contributors are acknowledged in the 
acknowledgments section rather than as co-authors.
\end{tcolorbox}
\end{textblock*}

\small  

\begin{abstract}
In this work, we survey the role of Graphics Processing Units (GPUs) in real-time systems. Although originally designed for parallel graphics workloads, GPUs are now widely used in time-critical applications such as machine learning, autonomous vehicles, and robotics due to their high computational throughput. Their massively parallel architecture enables significant acceleration of computationally intensive tasks, which can support stringent timing requirements. However, GPUs were not designed with timing predictability in mind, and their integration into real-time systems introduces several challenges, including limited and coarse-grained preemption, execution-time variability, and resource contention between concurrent tasks. These factors can lead to unpredictable delays and potential deadline violations. We review existing approaches that address these issues, including scheduling algorithms, resource-management techniques, and synchronization methods, and highlight open research directions aimed at improving GPU predictability and performance in real-time environments.
\end{abstract}

\section{Introduction}

GPUs are now dominant in embedded platforms, serving as high-performance accelerators for real-time tasks such as multi-object visual tracking and autonomous vehicles, where they handle computationally intensive operations \cite{liang2014efficient, bojarski2016end, lin2018architectural}. Despite their widespread use in these applications, GPUs face a significant challenge due to their lack of inherent real-time compatibility. Their asynchronous and non-preemptive nature can lead to delays for high-priority tasks, negatively impacting system predictability \cite{han2022microsecond, Basaran2012, hartmann2019gpuart, Zhou2015}.

Originally designed for graphics rendering, GPUs have evolved to support general-purpose computation, driving the increasing demand for parallel computing in diverse fields such as machine learning, robotics, and scientific computing \cite{dwarakinath2008fair, adriaens2012gpgpu, aguilera2014fair, capodieci2018deadline}. However, real-time workloads often require features such as preemption and multitasking, capabilities that GPUs have traditionally lacked. \newpage This limitation has spurred significant research into both software and hardware solutions to address challenges like variable kernel execution times, resource contention, and context-switching overhead. One promising solution is spatial multitasking, which improves resource sharing and provides substantial performance benefits over traditional methods for concurrent applications \cite{adriaens2012gpgpu}.

To enhance GPU resource management, various scheduling and multitasking approaches have been proposed. Despite progress in real-time systems research, several challenges remain, including limited preemption support, long blocking times, and the need for extensive code modifications. This work explores these developments, focusing on the performance and energy efficiency of GPU multitasking and scheduling techniques to identify future research directions for real-time and latency-sensitive environments \cite{wang2024unleashing}.

Based on our investigation, the use of GPUs in real-time systems can be categorized into two major areas: 1) scheduling algorithms that explicitly address real-time constraints, discussed in Section 4, and 2) scheduling algorithms that implicitly consider real-time constraints, thoroughly covered in Section 5.

The work is structured as follows: Section 2 covers GPU architecture and real-time systems. Section 3 discusses the obstacles in integrating GPUs into real-time systems. Sections 4 and 5 explore the two categories of GPU scheduling algorithms. Section 6 discusses future directions, and Section 7 provides the conclusion.
\section{Background}

Modern GPUs surpass recent multi-core CPUs in transistor count and have evolved into powerful general-purpose stream processors capable of handling complex floating-point computations. Beyond their traditional usage in improving 3D rendering, GPUs are now widely utilized for various general-purpose tasks, including scientific computations, data-intensive applications, and real-time systems \cite{dwarakinath2008fair}. Since this work deals with the use of GPUs in real-time systems, we focus on these main concepts in this section. 
\subsection{GPU Architecture and Execution Model}
We adopt Nvidia's terminology throughout this section. GPUs serve as massively parallel accelerators, built from thousands of lightweight cores. These cores are organized into units called Streaming Multiprocessors (SMs), each equipped with shared memory, registers, and local caches. A unified last-level cache is shared among all SMs to ensure efficient data access \cite{akbarzadeh2024h3dm, bashizade2019adaptive}.

GPU programs consist of two key components: host code, which runs sequentially on the CPU, and device code, or kernels, which execute in parallel on the GPU. Execution begins with the host CPU invoking CUDA APIs, which are processed by the user-space runtime system and GPU driver. These components convert high-level API calls into low-level commands and queue them via the stream manager, ensuring ordered execution within each stream \cite{kato2011operating}.

Data is typically transferred from CPU to GPU using cudaMemcpy, followed by kernel launch, where the host specifies parameters such as grid and block dimensions. If sufficient resources—such as registers, shared memory, or execution slots—are unavailable, the kernel launch is delayed until resources become available, as most GPUs do not support fine-grained preemption of running kernels \cite{li2018dynamic}.

Kernels follow the Single-Instruction-Multiple-Thread (SIMT) model, where multiple threads execute the same instruction across different data. Threads are grouped into Thread Blocks (TBs), which are dispatched independently to SMs. Within each TB, threads are organized into warps, groups of 32 threads that execute instructions in lockstep \cite{park2015chimera, li2018dynamic}. Each SM may execute multiple TBs simultaneously, but all threads in a block remain on the same SM. The number of blocks and threads is specified at launch, though actual concurrency is constrained by hardware limits \cite{wang2021balancing, amert2017gpu}.

Once a kernel is scheduled, its TBs are assigned to available SMs. Inside each SM, one or more warp schedulers select ready warps each cycle and issue their instructions to the SM’s execution units. By switching between warps whose operands are ready, the schedulers effectively hide stalls caused by memory latency or long-latency operations \cite{nvidia2014nvidia, li2018dynamic}. This fine-grained warp switching—enabled by having each warp’s register state already resident on the SM—maintains high throughput without requiring expensive context switching.

Beyond these low-level hardware scheduling mechanisms, the CUDA programming model also provides higher-level features to improve concurrency and resource utilization. For example, the Multi-Process Service (MPS) enables kernels from multiple processes to execute concurrently, allowing better sharing of GPU resources and reducing context-switching overheads \cite{nvidia2020mps}. CUDA APIs further provide synchronization primitives—such as events, stream operations, and barriers—to coordinate kernel execution across the host and device. While these API-level mechanisms manage execution ordering and inter-process sharing, the underlying hardware schedulers remain responsible for distributing TBs to SMs, ensuring efficient utilization of available compute units.
\subsection{Real-Time Systems}
Real-time systems are a category of embedded systems designed to respond to inputs within strict timing constraints. Their correctness depends not only on producing accurate results but also on meeting timing deadlines. These systems are essential in safety-critical domains such as automotive control, avionics, and medical devices, where missing a deadline can lead to severe consequences \cite{safari2022survey}. 

A deadline defines the maximum time a task may take to complete without harming the system. Real-time systems are categorized based on the criticality of meeting task deadlines. Hard real-time systems require absolute adherence to deadlines, as any missed deadline can lead to catastrophic outcomes, such as in flight control or medical monitoring. Firm real-time systems also enforce strict deadlines, but missing one results only in the output being discarded, not system failure, though overall efficiency is reduced. In contrast, soft real-time systems can tolerate occasional deadline misses without severe consequences, making them suitable for applications like multimedia streaming, where delays degrade performance but do not compromise functionality.

Real-time systems run various application models where tasks may be dependent or independent. Dependent tasks must follow precedence relations, often represented by Directed Acyclic Graphs (DAGs), with tasks as nodes and dependencies as edges. Tasks can be periodic, aperiodic, or sporadic based on their activation patterns. Periodic tasks generate jobs at regular, fixed intervals. Aperiodic tasks have jobs arriving irregularly without a fixed pattern. Sporadic tasks are a type of aperiodic task with a guaranteed minimum time between job arrivals.

A scheduling algorithm is a set of rules that determines the order in which tasks are executed in a real-time system. These algorithms can be classified into several main types: \textit{preemptive} algorithms allow a running task to be interrupted and replaced by another, while \textit{non-preemptive} algorithms let tasks run to completion without interruption, making scheduling decisions only after a task finishes. \textit{Static} algorithms base scheduling on fixed parameters assigned before task activation, whereas \textit{dynamic} algorithms adapt decisions based on changing runtime conditions. In \textit{offline} scheduling, the entire task set is planned before execution and stored for runtime use, while \textit{online} scheduling makes decisions dynamically as tasks arrive or complete. Lastly, \textit{optimal} algorithms aim to minimize a cost function or find a feasible schedule when possible, whereas \textit{heuristic} algorithms use approximate methods to approach optimal scheduling without guarantees.
\section{Challenges}
Integrating GPUs into real-time systems presents significant challenges due to their architectural differences from CPUs and the complexities of managing resources and synchronizing tasks. GPUs are designed for high throughput, making it difficult to meet the strict timing constraints of real-time applications. Key issues include implementation difficulties, variable kernel execution times, resource contention, power consumption, and context switching overhead.
\subsection{Difficulties in Implementation}
One of the primary challenges in adopting GPUs for real-time systems lies in the proprietary, closed-source nature of GPU drivers, which restricts access to low-level hardware controls essential for precise optimization and predictable behavior \cite{hong2017gpu}. This limitation complicates task scheduling, execution prioritization, and resource management on the GPU, leading to unpredictable latency and difficulties in meeting strict real-time deadlines \cite{yang2018avoiding}. Synchronization bottlenecks between CPU and GPU workloads, caused by resource contention and blocking, further threaten system stability in safety-critical applications such as autonomous driving \cite{yang2018avoiding}. Additionally, the CUDA concurrency model introduces complexity in coordinating heterogeneous CPU-GPU threads, often resulting in latency and inefficiencies when workloads are not well balanced \cite{yang2018avoiding}. 

Traditional CPU synchronization primitives such as mutex locks are inefficient on GPUs due to SIMT execution constraints, lack of blocking primitives, and warp divergence \cite{xiang2014warp}. Although some GPUs, such as AMD’s, offer hardware-level synchronization features like the Global Data Share (GDS), these are limited in scalability and support only a fixed number of synchronization barriers, restricting their applicability in complex real-time scenarios \cite{devices2012amd, xiang2014warp}. Together, these factors present significant obstacles to achieving efficient concurrency and synchronization on GPUs for real-time applications.
\subsection{Execution Time Variability and Scheduling Challenges in GPUs}

The inherent non-determinism of GPU operations challenges reliable real-time execution, largely due to largely non-preemptive CPU-GPU memory transfers causing unpredictable delays \cite{Basaran2012}. Hu et al. \cite{hu2016run} and Wang et al. \cite{wang2016simultaneous} show that sharing GPU resources among concurrent applications and running multiple kernels simultaneously increase execution time variability, proposing scheduling and resource-sharing techniques to improve fairness and reduce delays. Saha et al. \cite{saha2019stgm} introduce a space-time GPU management framework to enable concurrent task execution with timing analysis support. Olmedo et al. \cite{olmedo2020dissecting} analyze CUDA’s multi-level scheduler, revealing its role in execution variability during multitasking. Furthermore, multicore processor scheduling adds complexity due to shared access to caches, I/O, and buses, complicating Worst-Case Execution Time (WCET) estimation \cite{reghenzani2019real}. Dasari et al. \cite{dasari2013timing} emphasize that such resource contention significantly hinders precise timing analysis in multicore GPU environments. 
\subsection{Resource Contention and Underutilization}

The current state-of-the-art GPU management for real-time systems significantly underutilizes GPU resources due to the serialization of GPU kernel execution.
In real-time systems, resource underutilization and contention are key challenges when utilizing GPUs. These systems often struggle to fully exploit GPU resources due to a lack of proper scheduling mechanisms, kernel preemption, and workload management. Resource contention, where multiple tasks attempt to access GPU resources simultaneously, can significantly degrade performance, especially in systems with strict timing constraints. Concurrent kernel execution on GPUs can help mitigate the issue of resource underutilization by allowing multiple tasks to share GPU resources simultaneously. 

However, this can lead to resource contention if not managed effectively.
Research by Li et al. \cite{li2016efficient} proposes a framework that enhances concurrent kernel execution by optimizing thread-level parallelism and implementing cache bypassing, effectively reducing contention and improving resource utilization on GPUs. Another significant cause of resource contention is memory access conflicts.
To reduce contention and prevent real-time tasks from being swapped, the mlockall system call can be used to lock memory pages, as discussed by Reghenzani et al. \cite{reghenzani2019real}. The effects of shared cache contention and hierarchical memory access, along with the need to partition GPU resources among competing tasks, further complicate the goal of predictable execution. This issue is particularly challenging when high-priority tasks interfere with lower-priority ones, disrupting deterministic execution \cite{zou2023rtgpu}.
\subsection{Power Consumption}
Power management is a critical challenge when integrating GPUs into real-time systems, especially due to dynamic power fluctuations caused by varying workloads. These fluctuations can lead to inefficiencies and thermal constraints. While techniques like Dynamic Voltage and Frequency Scaling (DVFS) provide some mitigation, optimizing energy consumption remains complex given the diverse GPU architectures. Additional strategies, such as power capping and kernel-specific frequency tuning, are being explored to balance energy efficiency with real-time requirements. Schoonhoven et al. \cite{schoonhoven2022going} further address GPU energy optimization through auto-tuning.
\subsection{Context Switching Overhead}

Context switching on GPUs is typically avoided due to the considerable overhead associated with saving and restoring the entire GPU state during process or kernel transitions. Nonetheless, such switches are sometimes indispensable for real-time and high-priority workloads, as evidenced by studies from Park et al. \cite{park2015chimera}, Ukidave et al. \cite{ukidave2016mystic}, Wang et al. \cite{Wang2016, wang2017quality}, Wu et al. \cite{wu2017flep}, and Xu et al. \cite{xu2016warped}. To address the synchronization challenges arising from context switching, advanced mechanisms like HeteroSync’s atomic tree barrier have been developed, which coordinate thread blocks through global counter updates to prevent deadlocks \cite{mahapatra2020designing, sinclair2017heterosync}. Furthermore, timing interference caused by memory swapping and unpredictable page faults threatens real-time determinism; this can be mitigated by employing memory locking techniques such as the mlockall system call to ensure consistent task execution \cite{reghenzani2019real}.
\section{Explicitly Considering Real-Time Constraints}

In this section, the scheduling algorithms that are specifically designed to handle the stringent timing requirements
in GPUs are investigated. These schedulers incorporate features like task preemption, where higher-priority tasks can
interrupt and take precedence over lower-priority ones, ensuring that critical tasks meet their deadlines. By focusing
directly on meeting the exact timing constraints and managing task priorities explicitly, these scheduling methods
provide robust and reliable performance in demanding real-time applications.
In this section, we classify explicit real-time scheduling methods into two main categories: 1) Non-preemptive
scheduling methods and 2) Preemptive scheduling methods. Non-preemptive scheduling ensures that once a task starts
executing, it runs to completion without interruption, thus reducing the overhead caused by context switching and
allowing for more predictable execution times. In contrast, Preemptive scheduling allows tasks to be interrupted and
resumed later, ensuring higher-priority tasks can take precedence as needed, which enhances system responsiveness.
\subsection{Non-Preemptive Scheduling}

Non-preemptive scheduling runs tasks to completion without interruption, offering high predictability essential for real-time and embedded systems. It is divided into temporal and space-time scheduling. Temporal scheduling allocates fixed time slots to tasks, ensuring execution within predefined periods, which is vital for real-time guarantees. TimeGraph \cite{kato2011timegraph} and Gdev \cite{kato2012gdev} exemplify this by providing frameworks for controlled, timely GPU access. Space-time scheduling integrates temporal and spatial resource allocation to manage tasks with strict timing and resource demands; STGM \cite{saha2019stgm} balances these aspects to optimize real-time GPU performance. Table \ref{non} provides an overview of non-preemptive scheduling methods. A detailed discussion of these methods is provided in the book.

\afterpage{
\begin{table}[h]
\centering
\caption{Overview of Non-Preemptive Scheduling Methods}
\label{non}
\scriptsize
\setlength{\tabcolsep}{2pt} 
\renewcommand{\arraystretch}{1.2} 
\begin{tabular}{|c|p{0.08\linewidth}|p{0.115\linewidth}|p{0.075\linewidth}|p{0.143\linewidth}|p{0.10\linewidth}|c|p{0.1\linewidth}|p{0.1\linewidth}|p{0.112\linewidth}|}
\hline
\filcenter \textbf{Ref.} & \filcenter  \textbf{Realtime Model} & \filcenter \textbf{Multitasking Method} & \filcenter  \textbf{Task Model} & \textbf{Implementation Level} & \filcenter \textbf{Scheduling Priority} & \textbf{Fairness} & \filcenter  \textbf{Resource Utilization} & \filcenter \textbf{Energy Management} & \textbf{Performance Improvement} \\ \hline

~\cite{kato2011timegraph} & \filcenter  Soft & \filcenter  Temporal & \filcenter  Aperiodic & Software-based & \filcenter  Fixed and Dynamic(can be fixed manually) & \filcenter  - & \filcenter  \checkmark & \filcenter  - & \makebox[\linewidth][c]{\checkmark} \\ \hline

~\cite{kato2011rgem} & \filcenter Soft & \filcenter Temporal & Periodic and Aperiodic & Software-based & \filcenter Fixed & - & \filcenter \checkmark & \filcenter - & \makebox[\linewidth][c]{\checkmark} \\ \hline

~\cite{kato2012gdev} & \filcenter  Soft & \filcenter  Temporal & \filcenter Aperiodic & Software-based & \filcenter  Dynamic & \checkmark & \filcenter  \checkmark & \filcenter  - & \makebox[\linewidth][c]{\checkmark} \\ \hline

~\cite{elliott2012globally} & \filcenter  Soft & \filcenter  Temporal & \filcenter Periodic & Sofware-based & \filcenter Dynamic & - & \filcenter  \checkmark & \filcenter  - & \makebox[\linewidth][c]{\checkmark} \\ \hline 

~\cite{elliott2013gpusync} & \filcenter  Soft & \filcenter  Temporal & \filcenter Sporadic & Sofware-based & \filcenter Dynamic & - & \filcenter  \checkmark & \filcenter  - & \makebox[\linewidth][c]{\checkmark} \\ \hline 

~\cite{muyan2016multitasking} & \filcenter Soft & \filcenter  Temporal & \filcenter Aperiodic & Software-based & \filcenter  Dynamic & - & \filcenter  - & \filcenter  - & \makebox[\linewidth][c]{\checkmark} \\ \hline

~\cite{kim2018server} & \filcenter  Soft & \filcenter  Temporal & \filcenter Periodic and Sporadic & Sofware-based & \filcenter Fixed & - & \filcenter  \checkmark & \filcenter  - & \makebox[\linewidth][c]{\checkmark} \\ \hline \specialrule{.05em}{0.1em}{0.1em}

~\cite{joo2014resource} & \filcenter Soft & \filcenter Space-time & \filcenter Aperiodic & Software-based & \filcenter Fixed & - & \filcenter \checkmark & \filcenter \checkmark & \makebox[\linewidth][c]{\checkmark} \\ \hline

~\cite{kang2017priority} & \filcenter Soft & \filcenter Space-time & \filcenter Sporadic & Software-based & \filcenter Dynamic & - & \filcenter  \checkmark & \filcenter \checkmark &\makebox[\linewidth][c]{\checkmark} \\ \hline

~\cite{jain2019fractional} & \filcenter  Soft & \filcenter  Space-time & \filcenter Aperiodic & Software-based & \filcenter  No specific priority levels among the tasks & - & \filcenter  - & \filcenter - & \makebox[\linewidth][c]{\checkmark} \\ \hline

~\cite{saha2019stgm} & \filcenter Soft & \filcenter Space-time & \filcenter Periodic and Sporadic & Software-based & \filcenter Fixed & - & \filcenter \checkmark & \filcenter - & \makebox[\linewidth][c]{\checkmark} \\ \hline

~\cite{wang2019work} & \filcenter  Soft & \filcenter  Space-time & \filcenter Aperiodic & Software-based & \filcenter  No specific priority levels among the tasks & - & \filcenter  \checkmark & \filcenter \checkmark & \makebox[\linewidth][c]{\checkmark} \\ \hline

~\cite{sun2020real} & \filcenter  Soft & \filcenter  Space-time & \filcenter Aperiodic & Software-based & \filcenter Fixed & - & \filcenter  \checkmark & \filcenter  - & \makebox[\linewidth][c]{\checkmark} \\ \hline

~\cite{chen2020flexgpu} & \filcenter  Soft & \filcenter  Space-time & \filcenter Aperiodic & Software-based & \filcenter  Dynamic & - & \filcenter  \checkmark & \filcenter - & \makebox[\linewidth][c]{\checkmark} \\ \hline

~\cite{zhao2021exploiting} & \filcenter  Soft & \filcenter  Space-time & \filcenter Aperiodic & Software-based & \filcenter  No specific priority levels among the tasks & - & \filcenter  \checkmark & \filcenter - & \makebox[\linewidth][c]{\checkmark} \\ \hline

~\cite{wang2021balancing} & \filcenter  Soft & \filcenter  Space-time & \filcenter Periodic & Software-based & \filcenter  Dynamic & - & \filcenter  \checkmark & \filcenter \checkmark & \makebox[\linewidth][c]{\checkmark} \\ \hline

~\cite{zou2023rtgpu} & \filcenter  hard & \filcenter  Space-time & \filcenter Periodic & Software based & \filcenter  Fixed & - & \filcenter  \checkmark & \filcenter  - & \makebox[\linewidth][c]{\checkmark} \\ \hline

~\cite{bakita2023hardware} & \filcenter  Soft & \filcenter  Space-time & \filcenter Sporadic & Software and hardware based & \filcenter  Fixed & - & \filcenter  \checkmark & \filcenter  - & \makebox[\linewidth][c]{\checkmark} \\ \hline



\end{tabular}
\end{table}
}

\vspace{3cm}
\subsection{Preemptive Scheduling}
Preemptive scheduling on GPUs generally relies on coarse-grained interruption, where tasks are paused and resumed at a larger granularity, unless specialized hardware features, such as fine-grained preemption support, or compiler-based techniques, such as checkpointing, are employed to enable more precise control over task execution. This method plays a key role in systems where multiple tasks compete for shared resources, such as GPUs. It enhances responsiveness and ensures critical tasks are completed on time by managing execution across various levels, including software-based and compiler-based (Section \ref{P1}), OS-based (Section \ref{P2}), hardware-based (Section \ref{P3}), and hybrid techniques (Section \ref{P4}) that integrate methods from these approaches to improve flexibility and effectiveness.
\subsubsection{Software-based and Compiler-based preemptive scheduling}
\label{P1}
Relies on system software to manage task interruptions, providing flexibility for preemption without requiring specific hardware support. This approach is seen in frameworks like EffiSha introduced by Chen et al. \cite{chen2017effisha}, which enable efficient preemptive scheduling for GPU tasks. While it allows for easier integration with existing hardware, the reliance on software can introduce overhead due to frequent context switching, which may affect performance. Table \ref{tab:soft} provides an overview of preemptive scheduling methods (software-based and compiler-based). A detailed discussion of these methods is provided in the book.

\newcommand{\fontnormal}{\fontsize{8pt}{12pt}\selectfont}
\newcolumntype{P}[1]{>{\centering\arraybackslash}p{#1}}
\afterpage{
\begin{minipage}{\textwidth}
 \FloatBarrier
\begin{table}[H]
\centering
\caption{Overview of Preemptive Scheduling Methods (Software-based and Compiler-based)}
\label{tab:soft}
\scriptsize
 \setlength{\tabcolsep}{2pt}
\renewcommand{\arraystretch}{1.2}
\begin{tabular}
{|c|P{0.06\linewidth}|P{0.149\linewidth}|c|P{0.2\linewidth}|P{0.055\linewidth}|P{0.055\linewidth}|P{0.055\linewidth}|P{0.055\linewidth}|P{0.055\linewidth}|P{0.055\linewidth}|P{0.055\linewidth}|P{0.055\linewidth}|P{0.055\linewidth}|P{0.055\linewidth}|}
\hline
\multirow{2}{*}{\centering \textbf{Ref.}} & \multirow{2}{*}{\textbf{1}} & \multirow{2}{*}{\textbf{2}} & \multirow{2}{*}{\textbf{3}} & \multicolumn{1}{c|}{\textbf{Preemptive Techniques}} & \multicolumn{6}{c|}{\textbf{Metrics}} \\ \cline{5-15}

   &  & & & \textbf{4} & \textbf{5} & \textbf{6} & \textbf{7} & \textbf{8} & \textbf{9} & \textbf{10}   \\ \hline

\cite{calhoun2012preemption} & Soft & Aperiodic & Fixed & Compile Time Parser & \tickNo & \tickNo & \tickNo & - & \tickNo & - \\ \hline

\cite{Basaran2012} & Soft & Periodic & Fixed & \tickNo & \tickYes & \tickYes & \tickNo & - & \tickNo & -  \\ \hline

\cite{Zhong2013} & Soft & Aperiodic & Dynamic & Code (SASS-PTX) Modification & \tickYes & \tickYes & \tickYes & - & \tickNo & \tickYes \\ \hline

\cite{kim2013segment} & Hard & Sporadic & Fixed & \tickNo & \tickNo & \tickNo & \tickNo & - & \tickNo & - \\ \hline

\cite{pai2014preemptive} & Soft & Aperiodic & Dynamic & \tickNo & \tickYes & \tickYes & \tickYes & \tickYes & \tickYes & \tickYes \\ \hline

\cite{chattopadhyay2014limited} & Hard & Sporadic & Fixed & \tickNo & \tickNo & \tickNo & \tickYes & - & \tickYes & - \\ \hline 

\cite{park2015chimera} & Hard & Periodic & Dynamic & Using Compiler Analysis & \tickYes & \tickNo & \tickYes & - & \tickNo & \tickYes \\ \hline 

\cite{lee2016run} & Soft & Periodic & Fixed & Workload Splitter & \tickYes & \tickNo & \tickYes & - & \tickYes & \tickYes   \\ \hline 

\cite{chen2017effisha}, \cite{chen2016software} & Soft & Aperiodic & Dynamic & Code Transformation & \tickYes & \tickYes & \tickYes & - & \tickYes & \tickYes \\ \hline

\cite{garg2017gpuscheduler} & Soft & Aperiodic & Fixed & Adding Predefined Macros & \tickYes & \tickNo & \tickNo & - & \tickNo & - \\ \hline

\cite{sorensen2017cooperative} & Soft & Periodic & Dynamic & \tickNo & \tickYes & \tickNo & \tickNo & \tickYes & \tickYes & - \\ \hline

\cite{wu2017flep} & Soft & Aperiodic & Fixed & Code Transformation & \tickYes & \tickYes & \tickYes & \tickYes & \tickYes & \tickNo \\ \hline

\cite{jin2017preemption} & Soft & \textcolor{black}{Aperiodic} & Fixed & \tickNo & \tickYes & \tickNo & \tickYes & - & \tickYes & \tickYes \\ \hline

\cite{kim2017server} & Hard & Sporadic & Dynamic & \tickNo & \tickNo & \tickNo & \tickYes & - & \tickNo & - \\ \hline

\cite{garg2018share} & Soft & Aperiodic & Fixed & Code Modification & \tickYes & \tickYes & \tickNo & \tickYes & \tickYes & - \\ \hline

\cite{dong2018shared} & Hard & Sporadic & Fixed & \tickNo & \tickYes & \tickNo & \tickYes & - & \tickYes & - \\ \hline

\cite{bashizade2019adaptive} & Soft & Aperiodic & Dynamic & Kernel Modification & \tickYes & \tickYes & \tickYes & - & \tickYes & \tickYes \\ \hline

\cite{hartmann2019gpuart} & Hard & Periodic & Fixed & Code Modification & \tickYes & \tickNo & \tickYes & - & \tickYes & \tickNo \\ \hline

\cite{hosseinimotlagh2019thermal} & Hard & Sporadic & Fixed & \tickNo & \tickNo & \tickNo & \tickYes & - & \tickNo & - \\ \hline

\cite{eyraud2020algorithms} & Soft & Aperiodic & Fixed & \tickNo & \tickYes & \tickYes & \tickYes & - & \tickNo & - \\ \hline

\cite{Bai2020} & Soft & Aperiodic & Dynamic & Add Plugins to PyTorch Library & \tickNo & \tickYes & \tickYes & - & \tickYes & \tickYes \\ \hline

\cite{houssam2020hpc} & Soft & Sporadic & Fixed & \tickNo & \tickYes & \tickYes & \tickYes & \tickYes & \tickNo & - \\ \hline


\cite{yeung2021horus} & Soft & Aperiodic & Dynamic & \tickNo & \tickYes & \tickYes & \tickYes & \tickYes & \tickYes & \tickYes \\ \hline

\cite{roeder2021scheduling} & Hard & Aperiodic & Dynamic & \tickNo & \tickYes & \tickNo & \tickYes & - & \tickYes & - \\ \hline

\cite{ji2021collaborative} & Soft & Aperiodic & Dynamic & Kernel Transformer & \tickYes & \tickNo & \tickYes & - & \tickYes & \tickYes \\ \hline

\cite{yao2021wamp} & Soft & Periodic/Sporadic & Fixed & \tickNo & \tickYes & \tickYes & \tickYes & - & \tickYes & - \\ \hline

\cite{lopez2022flexsched} & Hard & Aperiodic & Dynamic & Kernel Transformation & \tickYes & \tickYes & \tickYes & \tickYes & \tickYes & \tickYes \\ \hline

\cite{chen2022pickyman} & Soft & Aperiodic & Dynamic & \tickNo & \tickNo & \tickYes & \tickYes & \tickYes & \tickYes & - \\ \hline

\cite{ayala2022new} & Soft & Aperiodic & Dynamic & Source Code Modification & \tickYes & \tickNo & \tickNo & - & \tickYes & - \\ \hline

\cite{bharmal2022real} & Soft & Periodic & Dynamic & Code Modification & \tickYes & \tickNo & \tickNo & - & \tickYes & - \\ \hline

\cite{xia2023towards} & Soft & Aperiodic & Dynamic & \tickNo & \tickYes & \tickYes & \tickYes & - & \tickYes & - \\ \hline

\cite{go2023selective} & Soft & Aperiodic & Dynamic & \tickNo & \tickYes & \tickYes & \tickYes & - & \tickNo & - \\ \hline

\cite{ng2023paella} & Soft & Aperiodic & Dynamic & CUDA Kernel Transformation & \tickYes & \tickYes & \tickYes & \tickYes & \tickYes & \tickYes \\ \hline

\cite{zhao2024cluster} & Soft & Aperiodic & Dynamic & \tickNo & \tickYes & \tickYes & \tickYes & - & \tickYes & \tickYes \\ \hline

\cite{han2024pantheon} & Soft & Aperiodic/Periodic & Dynamic & \tickNo & \tickYes & \tickYes & \tickNo & - & \tickYes & \tickYes \\ \hline

\end{tabular}
\end{table}

\vspace{-0.25cm}

\scriptsize
1: Real-Time Model,
2: Task Model,
3: Scheduling Priority
4: Compiler-based Technique,
5: Concurrency,
6: Profiling,
7: Cost/Time Estimation/Prediction,
8: Fairness,
9: Resource Management,
10: Throughput Improvement.
%

\FloatBarrier 
\end{minipage}
}
\subsubsection{OS-based preemptive scheduling}
\label{P2}
In OS-based preemptive scheduling, preemption is integrated directly into the operating system, which manages hardware resources more efficiently. Operating systems like VxWorks introduced by Barbalace et al. \cite{barbalace2008performance} and RTLinux by Ip et al. \cite{ip2001performance} have built-in support for preemptive scheduling, allowing them to handle both hard and soft real-time tasks with minimal delay. By coordinating hardware and software layers, OS-based approaches minimize latency, making them suitable for systems requiring real-time guarantees. Table \ref{tab:os} provides an overview of preemptive scheduling methods (OS-based). A detailed discussion of these methods is provided in the book.
\afterpage{
\begin{minipage}{\textwidth}
\FloatBarrier
\begin{table}[H]
\centering
\caption{Overview of Preemptive Scheduling Methods (OS-based)}
\label{tab:os}
\scriptsize
\setlength{\tabcolsep}{2pt}
\renewcommand{\arraystretch}{1.2}
\begin{tabular}
{|c|P{0.06\linewidth}|P{0.08\linewidth}|P{0.07\linewidth}|P{0.25\linewidth}|P{0.06\linewidth}|P{0.06\linewidth}|P{0.06\linewidth}|P{0.06\linewidth}|P{0.06\linewidth}|P{0.06\linewidth}|P{0.06\linewidth}|P{0.06\linewidth}|P{0.06\linewidth}|P{0.06\linewidth}|}
\hline
\multirow{2}{*}{\centering \textbf{Ref.}} & \multirow{2}{*}{\textbf{1}} & \multirow{2}{*}{\textbf{2}} & \multirow{2}{*}{\textbf{3}} & \multicolumn{1}{c|}{\textbf{Preemptive Techniques}} & \multicolumn{6}{c|}{\textbf{Metrics}} \\ \cline{5-15}

   &  & & & \textbf{4} & \textbf{5} & \textbf{6} & \textbf{7} & \textbf{8} & \textbf{9} & \textbf{10}  \\ \hline
   
\cite{takizawa2009checuda} & Soft & Aperiodic & Dynamic & Cuda Driver Modification & \tickNo & \tickNo & \tickNo & - & \tickYes & \tickYes \\ \hline

\cite{nukada2011nvcr} & Soft & Aperiodic & - & CUDA Driver Modification & \tickYes & \tickNo & \tickNo & - & \tickYes & - \\ \hline

\cite{capodieci2018deadline} & Soft & Sporadic & Dynamic & Driver Modification and Hypervisor & \tickYes & \tickNo & \tickNo & - & \tickYes & - \\ \hline

\cite{mizuno2018real} & Hard & Aperiodic & Dynamic & RT-Linux and Driver Modification & \tickYes & - & - & - & \tickNo & - \\ \hline

\cite{spliet2018case} & Hard & Periodic & Fixed & Driver Modification & \tickNo & \tickNo & \tickYes & \tickNo & \tickNo & \tickNo \\ \hline

\cite{eiling2022cricket} & Soft & Aperiodic & Dynamic & Virtualization & \tickYes & \tickNo & \tickNo & - & \tickYes & \tickNo \\ \hline

\end{tabular}
\end{table}

\vspace{-0.25cm}

\scriptsize
1: Real-Time Model,
2: Task Model,
3: Scheduling Priority
4: OS-based Technique,
5: Concurrency,
6: Profiling,
7: Cost/Time Estimation/Prediction,
8: Fairness,
9: Resource Management,
10: Throughput Improvement.
%

\FloatBarrier 
\end{minipage}
}
\subsubsection{Hardware-based preemptive scheduling}
\label{P3}
This scheduling relies on specific hardware mechanisms to support preemption. This reduces the overhead of task switching and enables faster responses to real-time events. Systems like those proposed by Lin et al. \cite{Lin2016} use hardware features for lightweight context switching, ensuring that high-priority tasks meet strict deadlines while maintaining system performance. Hardware-based solutions are especially useful in latency-sensitive environments where responsiveness is critical. Table \ref{hw} provides an overview of preemptive scheduling methods (hardware-based). A detailed discussion of these methods is provided in the book.

\afterpage{
\begin{minipage}{\textwidth}
\FloatBarrier

\begin{table}[H]
\centering
\caption{Overview of Preemptive Scheduling Methods (Hardware-based)}
\label{hw}
\scriptsize
\begin{tabular}
{|c|P{0.05\linewidth}|P{0.07\linewidth}|P{0.06\linewidth}|P{0.2\linewidth}|P{0.05\linewidth}|P{0.05\linewidth}|P{0.05\linewidth}|P{0.05\linewidth}|P{0.05\linewidth}|P{0.05\linewidth}|P{0.05\linewidth}|P{0.05\linewidth}|P{0.05\linewidth}|P{0.05\linewidth}|}
\hline
\multirow{2}{*}{\centering \textbf{Ref.}} & \multirow{2}{*}{\textbf{1}} & \multirow{2}{*}{\textbf{2}} & \multirow{2}{*}{\textbf{3}} & \textbf{Preemptive Techniques} & \multicolumn{6}{c|}{\textbf{Metrics}} \\ \cline{5-15}

   &  & & & \textbf{4} & \textbf{5} & \textbf{6} & \textbf{7} & \textbf{8} & \textbf{9} & \textbf{10}  \\ \hline
   
\cite{beaumont2019fine} & Soft & Aperiodic & Dynamic & \tickYes & \tickYes & \tickYes & \tickYes & - & \tickYes & \tickYes \\ \hline

\end{tabular}
\end{table}

\vspace{-0.25cm}

\scriptsize
1: Real-Time Model,
2: Task Model,
3: Scheduling Priority
4: Architectural Extensions,
5: Concurrency,
6: Profiling,
7: Cost/Time Estimation/Prediction,
8: Fairness,
9: Resource Management,
10: Throughput Improvement.
%

\FloatBarrier 
\end{minipage}
}
\subsubsection{Hybrid preemptive scheduling}
\label{P4}
Hybrid techniques combine methods from software, OS, and hardware-based approaches to harness the strengths of each. By integrating multiple methods, hybrid techniques aim to optimize preemption flexibility and reduce latency while balancing performance overhead. These techniques provide adaptable solutions for complex systems, achieving both responsiveness and efficiency in real-time environments. Table \ref{hybrid} provides an overview of preemptive scheduling methods (hybrid techniques). A detailed discussion of these methods is provided in the book.

\newcommand{\smallnormal}{\fontsize{4.9pt}{12pt}\selectfont}

\section{Implicit Real-Time Scheduling}

Implicit real-time scheduling refers to the use of conventional scheduling methods, which aren’t specifically designed for real-time applications but can still support real-time tasks under certain conditions. These methods typically rely on kernel concurrency and the inherent characteristics of the scheduler, such as fixed-priority or round-robin scheduling, or other simpler scheduling techniques, to meet deadlines indirectly. Although they are not tailored for the strict
requirements of real-time systems, in scenarios where timing demands are moderately flexible, these schedulers can
achieve satisfactory performance by using their natural scheduling behavior. This approach is generally less complex
and can be effective in systems where real-time support is beneficial but not critical. Table \ref{tab:merged_table} presents an overview of implicit real-time methods; for a detailed analysis of these works, readers are encouraged to consult the referenced book.


\afterpage{
\begin{minipage}{\textwidth}
\FloatBarrier

\begin{table}[H]
\centering
\caption{Overview of Preemptive Scheduling Methods (Hybrid Techniques)}
\label{hybrid}
\scriptsize
\begin{tabular}
{|c|c|P{0.075\linewidth}|c|P{0.13\linewidth}|P{0.11\linewidth}|P{0.08\linewidth}|c|c|c|c|c|c|c|c|c|}
\hline
\multirow{2}{*}{\centering \textbf{Ref.}} & \multirow{2}{*}{\textbf{1}} & \multirow{2}{*}{\textbf{2}} & \multirow{2}{*}{\textbf{3}} & \multicolumn{3}{c|}{\textbf{Preemptive Techniques}} & \multicolumn{6}{c|}{\textbf{Metrics}} \\ \cline{5-15}

   &  & & & \textbf{4} & \textbf{5} & \textbf{6} & \textbf{7} & \textbf{8} & \textbf{9} & \textbf{10} & \textbf{11} & \textbf{12}  \\ \hline

\cite{sajjapongse2013preemption} & Soft & Aperiodic & Fixed & \tickNo & Designing a Light-weight OS & \tickNo & \tickYes & \tickNo & \tickNo & - & \tickYes & - \\ \hline

\cite{tanasic2014enabling} & Soft & Aperiodic & Dynamic & \tickNo & SM Driver Modification & \tickYes & \tickYes & \tickNo & \tickYes & \tickYes & \tickYes & \tickNo \\ \hline

\cite{Zhou2015} & Soft & Aperiodic/ Periodic & Dynamic & Kernel Transformer & Driver Modification & \tickNo & \tickYes & \tickYes & \tickNo & \tickYes & \tickYes & - \\ \hline

\cite{zeno2016gpupio} & Soft & Aperiodic & Dynamic & Compiler and Runtime library & \tickNo & \tickYes & \tickYes & \tickYes & \tickYes & - & \tickYes & \tickYes \\ \hline

\cite{Lin2016} & Soft & Aperiodic & Fixed & \tickYes & \tickNo & \tickYes & \tickYes & \tickNo & \tickNo & - & \tickYes & - \\ \hline

\cite{li2018dynamic},\cite{li2018pep} & Soft & Aperiodic & Dynamic & \tickNo & \tickNo & \tickYes & \tickYes & \tickYes & \tickYes & - & \tickYes & - \\ \hline

\cite{lee2018gpu}, \cite{lee2020idempotence} & Soft & Aperiodic & Dynamic & Kernel Source Code Analysis & Driver Modification & \tickNo & \tickYes & \tickNo & \tickNo & \tickNo & \tickNo & - \\ \hline

\cite{long2020toward} & Soft & Aperiodic & Dynamic & \tickNo & OS-level scheduling & Device-level scheduling & \tickYes & \tickYes & \tickYes & \tickYes & \tickYes & \tickYes \\ \hline

\cite{calderon2022real} & Hard & Aperiodic & Dynamic & Source-to-source Transformation & PREEMPT-RT and Reverse Engineering of Driver & \tickNo & \tickNo & \tickYes & \tickYes & - & \tickYes & - \\ \hline

\cite{han2022microsecond} & Soft & Periodic & Dynamic & Code Transformation & Driver Modification & \tickNo & \tickYes & \tickYes & \tickYes & - & \tickYes & \tickYes \\ \hline

\cite{sun2022qos} & Soft & Aperiodic & Dynamic & Code Transformation & \tickNo & \tickYes & \tickYes & \tickYes & \tickYes & - & \tickYes & \tickYes \\ \hline 

\cite{wang2023secure}& Hard & Periodic & Dynamic & Code Transformation & Driver Modification & \tickNo & \tickYes & \tickYes & \tickYes & - & \tickYes & - \\ \hline

\cite{wang2024gcaps}, \cite{wang2024unleashing} & Soft & Sporadic & Fixed & Code Modification & Driver Modification & \tickNo & \tickNo & \tickNo & \tickYes & \tickYes & \tickYes & - \\ \hline

\end{tabular}
\end{table}

\vspace{-0.25cm}

\scriptsize
1: Real-Time Model,
2: Task Model,
3: Scheduling Priority
4: Compiler-based Technique,
5: OS-based Technique,
6: Architectural Extensions,
7: Concurrency,
8: Profiling,
9: Cost/Time Estimation/Prediction,
10: Fairness,
11: Resource Management,
12: Throughput Improvement.
%

\FloatBarrier 
\end{minipage}

\vspace{10pt}
}

\section{Future Directions and Applications}
The interplay of GPUs with real-time systems has driven notable advancements across diverse domains that require both high computational throughput and stringent latency guarantees. In robotics, real-time systems are foundational to essential functions such as localization, collision avoidance, and sensor data processing \cite{matsuoka2011robotics, laurent2014embedded}. Similarly, embedded systems exploit real-time capabilities to ensure dependable operation in automotive control and industrial automation applications \cite{lee2007handbook, verma2015slip}. The adoption of GPU-accelerated real-time processing in medical imaging has significantly improved the efficiency and responsiveness of data-intensive tasks \cite{eklund2013medical}. Additionally, telecommunications systems benefit from real-time frameworks that enhance data transmission and Quality-of-Service management \cite{atdelzater2000qos}. In cloud computing and the Internet of Things (IoT), real-time processing enables effective monitoring and control, supporting scalable and responsive infrastructures. Autonomous systems, including vehicles and drones, rely extensively on real-time decision-making mechanisms for navigation and obstacle avoidance. These topics will be explored in greater detail throughout this book.
\section{Conclusion}
GPUs have rapidly evolved into indispensable parallel computing platforms that significantly accelerate complex workloads in real-time and embedded systems, especially in critical domains such as autonomous vehicles and machine learning. Despite their computational power, inherent architectural features, such as non-preemptive execution, variable kernel latencies, and resource contention, present substantial challenges in meeting the strict timing and predictability demands of real-time applications. While recent research has made notable progress through advanced scheduling algorithms, multitasking frameworks, and resource management techniques to improve GPU responsiveness and utilization, important challenges remain. Future work must address the challenge of implementing efficient preemption mechanisms and reducing their associated overheads, while also adapting to emerging hardware features that increasingly support finer-grained preemption. Consequently, future efforts should focus on developing comprehensive and adaptive strategies that balance performance, predictability, and energy efficiency to enable the reliable integration of GPUs in safety-critical and latency-sensitive real-time systems.


\afterpage{
\begin{table}[H]
\centering
\caption{Overview of Implicit Real-Time Methods}
\label{tab:merged_table}
\fontsize{5}{7.35}\selectfont
\setlength{\tabcolsep}{3pt}
\renewcommand{\arraystretch}{1.2}
\begin{adjustbox}{totalheight=\textheight-5\baselineskip}
\centering
\begin{tabular}{|>{\centering\arraybackslash}m{0.05\linewidth}|
  >{\centering\arraybackslash}m{0.08\linewidth}|
  >{\centering\arraybackslash}m{0.08\linewidth}|
  >{\centering\arraybackslash}m{0.08\linewidth}|
  >{\centering\arraybackslash}m{0.085\linewidth}|
  >{\centering\arraybackslash}m{0.07\linewidth}|
  >{\centering\arraybackslash}c|
  >{\centering\arraybackslash}m{0.095\linewidth}|
  >{\centering\arraybackslash}m{0.08\linewidth}|
  >{\centering\arraybackslash}m{0.07\linewidth}|}

\hline
\textbf{Paper} & \textbf{Scheduling Granularity} & \textbf{HW/SW Implementation} & \textbf{Resource Sharing} & \textbf{Performance Improvement} & \textbf{Resource Utilization} & \textbf{Fairness} & \textbf{Power Consumption} & \textbf{Throughput Improvement} & \textbf{QoS Oriented}\\ \hline

~\cite{Fung2010} & Kernel & HW & Warp & \checkmark & \checkmark & - & - & - & $\times$ \\ \hline

~\cite{rossbach2011ptask} & Kernel & SW & GPU & \checkmark & \checkmark & \checkmark & - & \checkmark & $\times$ \\ \hline

~\cite{gregg2012resource} & TB & SW & GPU & \checkmark & \checkmark & - & - & \checkmark & $\times$ \\ \hline

~\cite{liang2014efficient} & TBs & HW &  SM & \checkmark & \checkmark & - & - & - & $\times$ \\ \hline

~\cite{aguilera2014qos} & Kernel & HW & SM & \checkmark & $\times$ & $\times$ & \checkmark & - & \checkmark \\ \hline

~\cite{jog2014application} & Kernel & HW & Memory & \checkmark & \checkmark & \checkmark & - & \checkmark & $\times$ \\ \hline

~\cite{aguilera2014fair} & TB & HW & GPU & \checkmark & \checkmark & \checkmark & - & \checkmark & $\times$ \\ \hline

~\cite{lee2014scheduling} & TBs & SW & Warps & \checkmark & \checkmark & - & - & - & $\times$ \\ \hline

~\cite{xu2016warped} & CTA & HW & SM & \checkmark & \checkmark & \checkmark & \checkmark & - & $\times$ \\ \hline

~\cite{suzuki2016towards} & Kernel & SW & GPU & \checkmark & \checkmark & - & - & - & $\times$ \\ \hline

~\cite{wang2016laperm} & TBs & HW & GPU & \checkmark & \checkmark & \checkmark & - & - & $\times$\\ \hline

~\cite{wang2016simultaneous} & TBs & HW & SM & \checkmark & \checkmark & \checkmark & - & \checkmark & $\times$ \\ \hline

~\cite{belviranli2016cumas} & Kernel & SW & GPU & \checkmark & \checkmark & $\times$ & - & \checkmark & $\times$ \\ \hline

~\cite{hu2016run} & Kernel & HW & SM & \checkmark & \checkmark & \checkmark  & - & - & $\times$ \\ \hline

~\cite{wang2017quality} & TB  & HW & Warp & \checkmark & \checkmark & $\times$ & \checkmark  & \checkmark & \checkmark \\ \hline

~\cite{hong2017fairgv} & Kernel & SW & Virtual GPUs & \checkmark & \checkmark & \checkmark & - & - & $\times$ \\ \hline

~\cite{liang2017exploring} & \color{black}TB & \color{black}HW & \color{black}Cache & \checkmark & \checkmark & - & - & \checkmark & $\times$ \\ \hline

~\cite{kim2018improving} & TBs & HW & Scheduling Unit, Execution Unit & \checkmark & \checkmark & \checkmark & - & \checkmark & $\times$ \\ \hline

~\cite{puthoor2018oversubscribed} & Kernel & HW & Warp & \checkmark & \checkmark & \checkmark & - & \checkmark & $\times$ \\ \hline

~\cite{jain2018dynamic} & Kernel & SW & GPU & \checkmark & \checkmark & - & - & \checkmark & $\times$ \\ \hline

~\cite{dai2018accelerate} & TB & HW & SM & \checkmark & \checkmark & \checkmark & \checkmark & - & $\times$ \\ \hline

~\cite{zhao2018classification} & Kernel & SW & SMs & \checkmark  &\checkmark &\checkmark &\checkmark & \checkmark & $\times$ \\ \hline

~\cite{xiao2018gandiva} & kernel & SW & GPU & \checkmark & \checkmark & \checkmark & - & - & $\times$ \\ \hline

~\cite{zhao2018efficient} & Kernel & SW & Virtual GPUs & \checkmark & \checkmark & - & - & \checkmark & \checkmark \\ \hline

~\cite{allen2019slate} & Kernel & SW & SM & \checkmark & \checkmark & - & - & \checkmark & $\times$ \\ \hline

~\cite{sun2019smqos} & Kernel & SW and HW & SMs & \checkmark & \checkmark & - & \checkmark & \checkmark & \checkmark \\ \hline

~\cite{shekofteh2019ccuda} & Kernel & SW & SM & \checkmark & \checkmark & - & - & - & $\times$ \\ \hline

~\cite{cruz2019maximizing} & Kernel & SW & GPU & \checkmark & \checkmark & \checkmark & - & \checkmark & $\times$ \\ \hline

~\cite{yu2019salus} & Iteration & HW & GPU & \checkmark & \checkmark & \checkmark & - & $\times$ & $\times$ \\ \hline

~\cite{tasoulas2019improving} & Kernel & HW & SM & \checkmark & \checkmark  & - & \checkmark & \checkmark & $\times$ \\ \hline

~\cite{lopez2020heuristics} & Kernel & SW & GPU & \checkmark  & \checkmark  & -  & - & - & $\times$  \\ \hline

~\cite{zhao2020hsm} & Kernel & HW & SM &  \checkmark  &\checkmark &\checkmark & - & \checkmark & \checkmark \\ \hline

~\cite{chen2021gemini} & Kernel & SW & GPU & \checkmark & \checkmark & \checkmark & - & \checkmark & $\times$ \\ \hline

~\cite{wu2021switchflow} & Kernel &  SW & GPUs, CPUs & \checkmark & \checkmark & - & - & \checkmark & $\times$ \\ \hline

~\cite{di2021tlb} & Kernel & HW & SM & \checkmark & \checkmark & - & - & \checkmark & $\times$\\ \hline

~\cite{chen2021smcompactor} & TBs & SW & SM & \checkmark & \checkmark & - & - & - & $\times$ \\ \hline

~\cite{parravicini2021dag} & Kernel & SW & SM & \checkmark & \checkmark & - & - & \checkmark & $\times$ \\ \hline  

~\cite{han2021flare} & TB & SW & SM & \checkmark & \checkmark & - & - & \checkmark & \checkmark \\ \hline

~\cite{weng2022raise} & Kernel & SW & SM & \checkmark & \checkmark & - & - & \checkmark & $\times$ \\ \hline

~\cite{wang2023dynamic} & Kernel &  HW  & SM & \checkmark  & \checkmark  &\checkmark  &\checkmark & - & \checkmark \\ \hline

~\cite{strati2024orion} & Kernel & SW & SM & \checkmark & \checkmark & - & - & \checkmark & $\times$ \\ \hline

\end{tabular}
\end{adjustbox}
\end{table}
}

\bibliographystyle{unsrt}
\bibliography{sample-base}

\end{document}